# Mentor Protégé Matchmaking Framework for IT Entrepreneurs: A Qualitative Study

Jamshed Memon[1], Mohd Zaidi Abd Rozan[2], Kamariah Ismail[2], Agha Jahanzeb[3]

[1]Department of Computer Science, Barrett Hodgson University, Karachi, Pakistan.
[2]UTM Technology Entrepreneurship Centre (UTMTEC),UTM Skudai, 81310 Johor, Malaysia
[3]Department of Finance, Sukkur Institute of Business Administration, Sukkur, Pakistan

**ABSTRACT**

This study proposes a framework for matchmaking a mentor and mentee by exploring intentions of entrepreneurs towards mentoring. The framework based on Theory of Planned Behaviour, Institutional Theory and Social Exchange Theory were applied to explore factors that influenced intentions of IT-Entrepreneurs in Malaysia towards mentoring. It was applied at three different phases to understand how selection of mentor changes from one phase to another, and identify type of mentoring support that entrepreneurs need during each phase. Qualitative data were collected from entrepreneurs through semi-structured interviews and analysed using NVivo10. Findings showed that entrepreneurs need different mentoring support and skills when they go through each phase of their entrepreneurial career. The conception phase is the most important phase because entrepreneurs are still in the process of developing their product and want a mentor to help them to evaluate the viability of their product. Entrepreneurs at this phase are also more concerned about the surface level characteristics of mentor such as age, gender, race and language than the other phases. At the start-up phase, entrepreneurs want a mentor who can provide them networking support and help them in decision making. Besides that, the IT entrepreneurs think that mentoring services provided in the incubators are very generic and not helpful. On the contrary, entrepreneurs at the growth phase want a mentor whom they see as a role model and is more experienced and successful. Entrepreneurs at this phase do not care much about factors such as gender, race and religion of the mentor.

.
## 1. Introduction

Entrepreneurship is the attitude and process to create and improve economic activity by creativity, innovation with rigorous management in a new or an existing organisation (European Commission, 2006). Entrepreneur needs support while growing up and a mentor can provide that support during the difficult times(Eesley and Wang, 2014). Match making between entrepreneur and mentor is very important because during the review of mentoring within different fields like education and business, Ehrich (2004) found that experience or matchmaking was one of the biggest problem in a mentoring relationship. Problem with the existing studies on mentor-protégé relationship is that respondents in these researches were already in a mentor-protégé relationship (Bell and Treleaven, 2010; Cain, 2009; Fruchter and Lewis, 2003; Harris, Freeman, and Aerni, 2009; Haynes and Petrosko, 2009). Therefore, this research explores the factors which influence the protégé's intentions towards the mentoring, whereas respondents are not in the relationship as yet, but are seeking a mentor.

## 2. Problem Background

The start-up survival rate is very low as research by SBA (Small Businesses Association) of the USA foundthat only 50% of business survive till 5 years (Cook, Campbell, and Kopp, 2015). According to SBA office of Advocacy (2012) report on small businesses, there are more companies closing down in the USA than there are new companies opening up. Figure 1.1 shows the new employer firms births and deaths over 10 years. Statistics show that during 2008-09 approximately 0.52 million new companies were born while about 0.68 million companies closed down. Mentoring is suggested to be an option to increase the survival rate of start-ups (Sullivan, 2000). Good mentoring can increase the survival rate of start-ups, because entrepreneurs usually lack competency and experience due to which it is hard for them to find a profitable niche for their businesses (St-Jean and Audet, 2012).

## 3. Conceptual Theoretical Framework

One of the research areas in entrepreneurial mentoring is that of matchmaking of mentor and protégé. Entrepreneurs need support whilst starting up and a mentor can provide that support during the difficult times. Existing research mainly focuses on studying/examining mentoring relationships where the mentor and protégé are already in a relationship (Allen, 2004; Baranik, Roling, and Eby, 2010; Eller, Lev, and Feurer, 2014).

Mentoring is relevant to multiple theories. These include the Theory of Planned Behaviour (Allen, 2003; Leck, Orser, & Riding, 2009), Commitment (Donaldson, Ensher, and Grant-Vallone, 2000; Joiner, Garreffa, Bartram, and Management., 2004; Ragins, 1997). Leadership (Scandura and Williams, 2004) and Social exchange theory (Eby, Buits, Lockwood, and Simon, 2004; Ensher, Thomas, and Murphy, 2001). In addition, diversity, including gender has also been in focus in the mentoring literature (Leck et al., 2009; Ragins, 1997; Thomas, 1990; Turban et al., 2002). Diversity is often discussed as surface level diversity or deep level diversity (Harrison, Price, and Bell, 1998; Price, Harrison, Gavin, and Florey, 2002). Surface level diversity consists of visible attributes which are readily detectable such as age and gender. Memon Jamshed, M.Z.A Rozan, Mueen Uddin, Asadullah Shah (2013) describe surface level diversity as objective qualities of a mentor and deep level diversity as subjective. Deep level diversity is defined as differences among attitude, beliefs and values (Harrison et al., 1998; Price et al., 2002). As both level of diversity affect the relationship, this study will explore the issue concerning surface-level (Age, Gender, Race) and deep level diversity (Personality, beliefs and Values) which affects the intentions of entrepreneurs towards mentoring.

The proposed framework consists Theory of Planned Behaviour (Ajzen, 1991), social exchange theory (Fiske, 1991; Fiske, 1992). Fiske's Social Exchange Theory provides an alternative explanation for the mentoring process (Rutti, Helms, & Rose, 2013). The proposed framework is an extension of the framework proposed by Memon et al, (2014). The reason, entrepreneurs wanting to be mentored and what support they seek from a mentor (Leck et al., 2009; St-Jean, 2012)can be explained. Subjective norms will help explain who influenced the intentions of an entrepreneur towards mentoring.

Surface-Level Diversity and Deep Level diversity concepts have been adopted from SET (Rutti et al., 2013). Both will be used to assess the qualities of a mentor. Table 1 outlines the constructs being used in this theory together with the literature about the origin of these constructs. While the figure.1 shows the proposed research framework and the background factors that influence the constructs.

Table 1: Theory Formulation (Memon et al,, 2014)

| Construct | Definition | Instrument |
|---|---|---|
| (Attitude) Mentoring Functions | Persons's +ve or –ve attitude towards performing a behaviour | (Kram, 1985; Scandura, 1998) |
| Subjective Norms | Perceived social pressure to perform or not to perform the behaviour | TPB/TRA |
| Past Experience | Influence of past behaviour/experience on intentions. | (Carr and Sequeira, 2007) |
| Surface-Level Diversity | Exploring the Readily detectable features Such as Age, Gender | (Fiske, 1991 Fiske, 1992) |
| Deep-Level Diversity | Exploring the Personality beliefs, values | (Fiske, 1991; Fiske, 1992) |

Figure 1: Theoretical Framework (Memon et al., 2014)

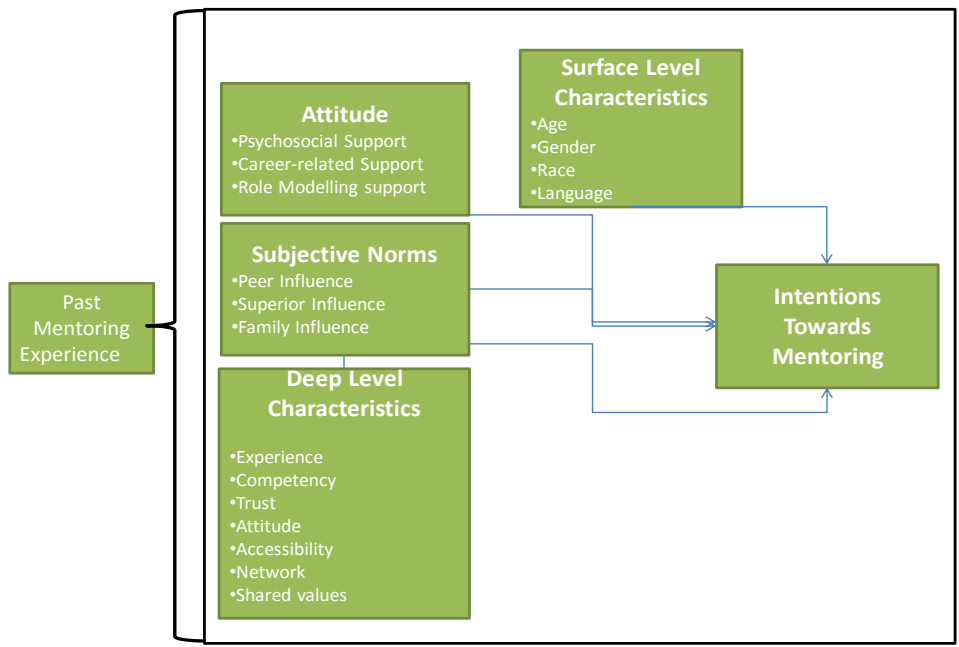

## 4. Research Methodology

Qualitative/Interpretive research paradigm was chosen for this research. The reason to opt for the qualitative research mode is to create an empirically valid framework, which can be used to measure the mentoring intentions of the entrepreneurs. Qualitative research helps in formation of concepts and the interrelationship between those concepts. On the other hand, quantitative study is more concerned about discovery of facts about the phenomenon and testing hypothesis on respondents chosen from a sample of population in order to achieve the statistical generalisation (Hoepfl, 1997; Meredith, 1998). Therefore, it was important to create an empirically valid framework before statistical generalisation can be achieved.

### 4.1 Interview Protocol

A Semi-Structured interview was conducted based on the constructs of the theory of planned behaviour and Social Exchange Theory. Interview protocol started with the introduction about the researcher. Respondent was told that who interviewer was, what was the qualification of researcher, research purpose and what the researcher wanted to ask from the respondent (Hill et al., 2005). Seven structured questions. The rest of the questions were non-structured and varied from interview to interview, depending on saturation of data. Louise Barriball and While (1994) was followed for conducting face to face interviews.

### 4.2 Interviews

Interviewees were categorised into three different phases to establish validity of data and to explore the type of mentor, entrepreneurs want at each stage of their entrepreneurial career.

This phase was named as conception phase, because students were still developing their products and had not yet started the business. Few of the respondents had secured RM 160,000 funding from the private companies. The research findings presented in this phase are based on semi-structured interviews of 17 nascent ICT Entrepreneurs. The data in this phase were collected from entrepreneurs who were still at an early stage of their career and had not developed their product yet.

During the second phase, researcher interviewed eight entrepreneurs who had already started their business and were generating revenue up to RM 300,000. SMECorp 2014 Classification of small and medium enterprise was used to classify the entrepreneurs. According to SMECorp classification 2014, companies with revenue up to RM 300,000 and less than 5 employees were classified as Micro organisations.

During the third phase, which is also called growth phase (Sullivan, 2000), five entrepreneurs were interviewed. Three of them were face to face and two were conducted using email because of non-availability of entrepreneurs for face to face interviews. All of the five respondents were male. Organisations were classified as small business organisations according to SMECorp classification of Small and Medium Enterprises. Organisations had annual turnover of between RM 300,000 to

RM 3,000,000 and had 5 to 20 employees. Companies established were 2 to 4 years old. Entrepreneurs were interviewed face to face and via email. Subsequent emails were sent for clarification and probing of the answers.

**4.3 Directed Content Analysis**

Sometimes a research exists about a phenomenon which is incomplete or it needs further description. Directed content analysis method is used when researcher wants to validate existing theory or improve/extend the theory (Hsieh and Shannon, 2005). The broad concepts or constructs, mentioned in conceptual framework were refined into themes and subthemes, based on a pattern in the data and from concepts which frequently occurred. Results were consistent across all interviews and thus are organized around 6 main topic areas that were used to create a Theoretical framework, namely, 1) Attitude, 2) Subjective Norm, 3) Surface Level Characteristics, 4) Deep Level Characteristics and 5) Past mentoring experience. A New Concept named 6) Environment emerged during analysis of data which were not part of the conceptual framework. This new concept was adopted from institutional theory. During the analysis of the data, it was observed that it was not only influence of individuals like, family, friends and teachers that influenced the intentions of entrepreneurs towards mentoring, but it was also normative and coercive influence. Indicator of normative influence measured if entrepreneur learned about mentoring during education or learned it through online media like internet and television (Honig and Karlsson, 2004). While indicators of coercive influence measured if the entrepreneur's intentions towards mentoring were influenced after attending an incubator program or seminar or workshop.

**5. Empirical Framework**

The researcher came up with a theoretical framework based on the empirical data. The initial theoretical framework (see figure 1) was used as a lens to code the data into themes and sub themes during three phases in order to explore the factors that influence the entrepreneurs' intentions towards the mentoring at the each phase. The frequency of factors reported by entrepreneurs varied from one phase to another. Table 2 shows the number of times each theme and sub theme were reported during conception, start-up and growth phase. While the figure 2 shows the revised theoretical framework, which was developed using the empirical data on entrepreneurs' intentions towards mentoring.

**Table 2:** Summary of construct, themes and sub themes at conception, start up and growth phase

| Constructs | Key Themes | Sub Themes | Conception Phase N = 17 | Start-up Phase N = 8 | Growth Phase N = 5 |
|---|---|---|---|---|---|
| Attitudinal Belief | Entrepreneurial Support | Information Support | 14 | 2 | 1 |
| | | Networking support | 7 | 2 | 2 |
| | | Knowledge Support | 5 | 3 | 3 |
| | | Problem Solving | 4 | 3 | 0 |
| | Psychological Support | Confidence | 12 | 5 | 1 |
| | | Decision Making | 6 | 5 | 2 |
| | Role Model Support | Reflection | 5 | 3 | 4 |
| Subjective Norms | Peer Influence | | 6 | 3 | 3 |
| | Superior Influence | | 7 | 2 | 2 |
| | Family Influence | | 1 | 0 | 0 |
| Environment | Normative Influence | | 8 | 2 | 2 |
| | Coercive Influence | | 2 | 2 | 2 |
| Surface Level Characteristics | Age | | 12 | 4 | 2 |
| | Gender | | 5 | 1 | 2 |
| | Race | | 7 | 2 | 1 |
| | Language | | 6 | 2 | 1 |
| | Geographical Distance | | 9 | 4 | 2 |
| Deep Level Characteristics | Accessibility | | 15 | 8 | 5 |
| | Personality | | 13 | 6 | 4 |
| | Skills | | 12 | 7 | 4 |
| | Experience | | 10 | 7 | 5 |
| | Religion | | 9 | 2 | 0 |
| | Partnership | | 0 | 4 | 3 |
| Past Experience | | | 5 | 7 | 4 |

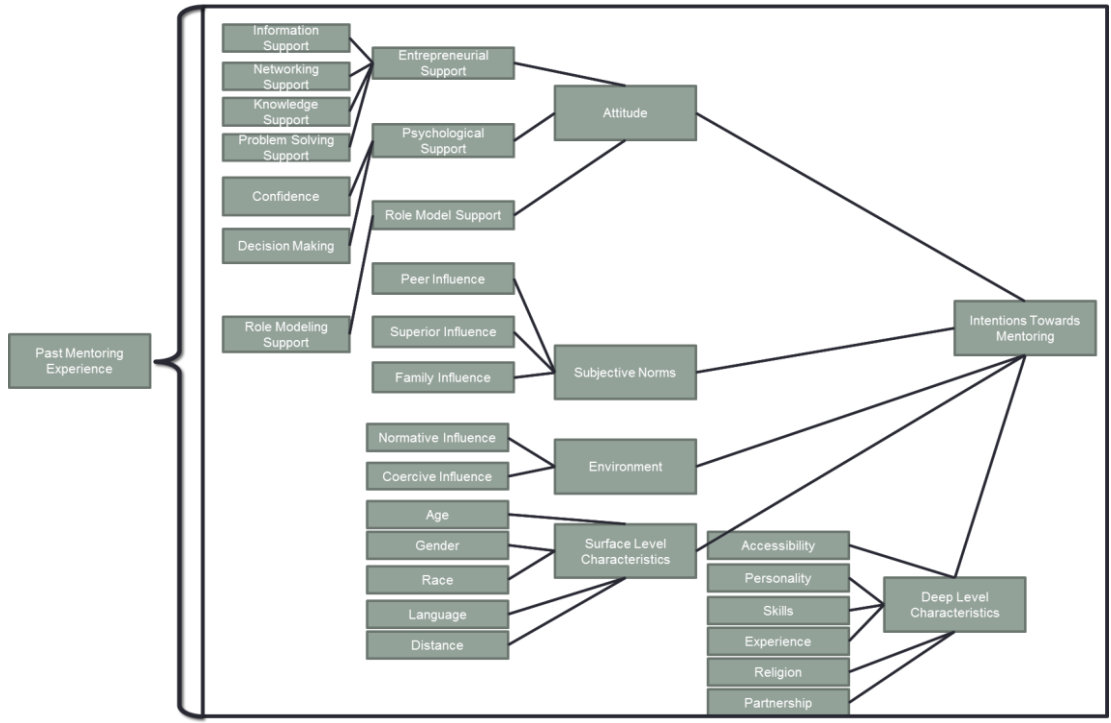

**Figure 2**: Revised Theoretical Framework of mentor-protégé matchmaking

## 6. Findings

Each theme along with the sub theme is discussed in detail during this section. Naming convention was used to represent the respondents e.g. Entrepreneurs E1P1 represents the entrepreneur number 1 at conception phase. Similarly entrepreneur E3P2 identifies the entrepreneur 3 at phase 2 i.e. start-up phase.

### 7.1 Attitudinal Belief

The first construct, attitudinal belief of theory of planned behaviour (Ajzen, 1991) refers to individuals confidence that a mentor can provide support which is needed during the different phases of entrepreneurial career. This construct was measured through mentoring support that entrepreneurs want from a mentor. Three factors emerged from Mentoring namely Entrepreneurial, Psychological and role model support (Leck et al., 2009; St-Jean, 2012). Career related support factor was changed to entrepreneurial support because, literature suggests that career related support factor has been used in organizational and career related mentoring studies(Allen, 2003; Kram, 1985; Leck et al., 2009; Rutti et al., 2013).

Attitude construct shows that entrepreneurs at conception phase are more inclined to get entrepreneurial support from mentor. Mentor is being seen as someone who is more skilled and knowledgeable. Literature also suggests that entrepreneurs received more entrepreneurial support than psychological or behavioural support (Kyrgidou and Petridou, 2013; Sarri, 2011). Information support was the most prominent sub theme during this phase. It may be because entrepreneurs were still at conception phase. It can be, because the entrepreneurs were computer graduates, having less knowledge about marketing and sales.

Entrepreneurs at start-up phase were more interested in psychological support as compared to nascent entrepreneurs at conception phase who were more inclined to get entrepreneurial support. Emerging themes show that start-up entrepreneurs want their mentors to help them in decision making and also boost their confidence during the course of relationship.

During the growth phase, it was observed that the trend towards networking and role model support was on the rise while information and knowledge support were on decline. It may be because entrepreneurs wanted their mentors to help them in getting projects and recommend them to the clients. It was also observed that 4 of 5 entrepreneurs at growth stage had past mentoring experience and 2 of them had more than one mentors. One entrepreneur reported that his relationship with the mentor couldn't work because the mentor was provided by the incubator and was not the type of mentor he wanted.

### 6.1.1 Entrepreneurial support

Figure 2 shows that entrepreneurial support consists of 4 sub factors: information support, networking support, knowledge support and networking support. Entrepreneurial support is form of direct support that helps entrepreneurs in building the company, managing the team, cash flow, helps in marketing and sales and helps in solving problems that entrepreneurs face during different phases of their career.

St-Jean and Audet (2012)re-defined career support as entrepreneurial support because career support was empirically being used in organizational mentoring context (Botha, 2014; Horvath et al., 2008; Kram, 1983; Leck et al., 2009). Psychosocial support was re-defined as psychological support because socialisation and networking support which is empirically part of psychosocial support (Allen, 2003; Hamlin and Sage, 2011; Kram, 1983; Santos and Reigadas, 2002) was made part of entrepreneurial support.

### 6.1.1.1 Information Support

Entrepreneurs were looking for information support because being an entrepreneur it was important to have information about every aspect of business and not just technical knowledge which they had already e.g. Entrepreneurs at conception phase were more interested in information support on sales and marketing. E3P1 reported that information support would help the entrepreneurs in avoiding mistakes and it will help them in devising a better sales and marketing strategy.

*"To build a company we must know what we want to do. With the mentor I think… we can avoid doing mistakes. We will be able to do it quick and save our time and spend that time on product and services. I think because I am from a technical background so I need more mentoring on marketing and sales because I cannot do both because I am being technical and don't have knowledge of both so I need mentoring"*E3P1.

Entrepreneur E5P1 and E17P1 also supported it and they believed that being from computer science background, they did not have enough knowledge about business and sales. On the other hand, the entrepreneurs at start-up phase needed more support on building the company and the projects. Entrepreneur E2P2 reported that mentor can help in providing information on managing the project and the company, which otherwise may take more time.

*"As a start-up entrepreneur I want to be mentored because I do not have experience in doing business. Specially how to manage your company. How to manage your projects. I can do it on my own, but it takes longer time to be good entrepreneur without so much risk. With mentoring you can lower the risk of failure"* E2P2.

Only one entrepreneur at growth phase reported about the information support. Entrepreneur E4P3 reported that mentor can keep entrepreneur on track, because sometimes entrepreneurs at growth stage get lost. Therefore it is good to have a mentor who can be more like an advisor.

*"Because I want to get guidance on solving problems. In the course of entrepreneurship, I will encounter all sorts of problems. Sometime I am at lost on what to do. It will be good to have a mentor to give me advises from time to time"*E4P3.

Literature suggests that entrepreneurs want their mentors to provide them knowledge and Information support (Radu Lefebvre and Redien-Collot, 2013; St-Jean and Audet, 2012, 2013). According to Radu Lefebvre and Redien-Collot (2013), nascent entrepreneurs need more information support at the beginning of the relationship when entrepreneurs are still developing their products while they will need more psychological support when they launch their product. This research also confirms that entrepreneurs do need information support at the beginning while more psychological support once they launch their product But this research also highlights the type of information support that ICT entrepreneurs need at conception, start-up and growth phase, which the literature lacks.

**6.1.1.2 Networking support**

Entrepreneurs wanted to use the contacts of the mentors to get into the market. Entrepreneur E2P1 at the conception phase reported that because of the lack of experience it will be difficult to get into the market Therefore, mentor can help the entrepreneur by connecting him with others. Entrepreneur E15P1 wanted to use mentor's contacts to get clients for the web development business.

*"Mentor should help me improve my self-esteem and teach us how to make good communication with other people and also introduce me to other people in the field"* E15P1.

At the start-up phase entrepreneurs were more specific about the type of people they wanted their mentor to help them connecting with. Entrepreneur E2P2 reported that, mentor could help in connecting with the companies which wanted to use social media marketing for their products.

*"I want my mentor to help me getting more clients. May be other companies who want to use internet and social media for their products. I believe that in Malaysia there are still a lot of companies which don't use internet for selling and marketing their products. So I want someone who can get me projects and more business"* E2P2.

Entrepreneurs at growth phase wanted to use the contacts of the mentor to expand their business and get funds. Entrepreneur E3P3 reported that his company was doing well and getting good clients, but growth was slow. Therefore, a mentor could help his company get funds which will help his company expand.

*"I think I will have access to funds because mentor will have network so I can get reference to other firms who can help me to grow my business. It's like I will have access to market, I will get more knowledge like how I get more resources and grow my company. So I think this mentor will provide help"* E3P3.

Literature suggests that nascent entrepreneurs want their mentors to provide them networking support, where they can integrate with the entrepreneurial community. They want their mentors to provide them contact of entrepreneurs who can help them achieve their objectives (Radu Lefebvre and Redien-Collot, 2013). According to St-Jean and Audet (2012), mentors should facilitate the integration of protégé in the business community by representing the protégé to the business contacts who may be of the need in future. According to McGregor and Tweed (2002), networking support can help the women entrepreneurs in their business growth, because they will be able to overcome the business isolation by connecting with the other women in the business. Eby and Lockwood (2005)found that in organizations, employees also wanted to get networking support from their mentors to remain marketable both within, and outside of their organisations.

### 6.1.1.3 Knowledge support

Entrepreneurs who reported about the knowledge support were more explicit and clear about the support they wanted to get from mentor e.g. at the conception phase novice entrepreneurs were looking for support like, pitching, preparation of business plan and securing the pre-seed funding. Entrepreneur E16P1 reported that mentoring would help in preparing a good business plan to secure the funding. E10P1 reported that mentor could help in bringing the software product that entrepreneur was developing into the market, because a mentor would know the right customer for the product.

*"I need a mentor to guide me about real world of entrepreneurship and about the market and give me the real situation in business area. Like my product is about a software and app about printing service so I want to know from mentor that how can I bring this product into the market and how my product can penetrate into the market"* E10P1.

At the start-up phase entrepreneurs want a mentor who could help the entrepreneurs in dealing with the customers. Entrepreneur E3P2 reported that it was difficult to handle the customers for them, because of lack of management experience. Entrepreneur E6P2 also reported that a mentor can guide us on how to bring more clients in order to grow and expand the business.

*"I will be able to get more projects because at the moment, company is new and I have to build a portfolio to show other companies that I can offer better solution to them. So a mentor can guide me that how should I deal with clients, about billing and pricing and about breaking even"* E6P2.

At the growth phase entrepreneurs want a mentor who can help in building leadership skills, entrepreneur E5P3 reported that team building was more important, because apart from timely completion of the project, customer satisfaction was also important. Therefore a mentor can help the entrepreneur in team building.

*"A mentor can provide skills like how to manage the team and projects. As an entrepreneur I don't have much knowledge about team management so I need some leadership skills to manage the team. Project management is not just about timely completion of project but it is also about customer satisfaction so I want both of the skills"* E5P3.

Knowledge acquisition or knowledge support has been reported by (Leck et al., 2009) as one of the support that protégés reported to have received during mentoring. Demick and Reilly (2000) found that managing director of a company went through mentoring reported that mentoring helped in team building and analysis of new marketing opportunities. Bosi, Pichetti, and Tudor, (2012) claimed that mentoring develops the knowledge skills of novice entrepreneurs. (Bisk) 2002 identified an interesting aspect that mentees do not look for an industry specific knowledge, despite the fact that organizations which provide mentoring services often focus on specific fields. This claim was confirmed during the data analysis that entrepreneurs wanted knowledge and support which was not industry specific, but Dalley and Hamilton (2000) stressed that successful development of knowledge which leads to business development depends on mentor's credibility and compatibility with the entrepreneur or protégé.

### 6.1.1.4 Problem Solving Support

Problem solving support is related to specific problems that entrepreneurs face. Entrepreneurs want advice of mentor on specific problems. Entrepreneur E3P1 at the conception phase reported that he wanted the mentor to help him in choosing a marketing strategy that how entrepreneur can reach out to the customers. Entrepreneur wanted to know that whether social media marketing should be used or if there was any other way to market the product. Entrepreneur also reported another problem about the funding and wanted to know whether he should go for the seed funding from the government or the private investor. Entrepreneur wanted the mentor to guide him about the pros and cons of government funding and private angel investment.

*"....Let's say we go to have funding from outside. The Investor would also want certain percentage in the company. So I and my mentor can decide whether we should go for funding or not or maybe we can just rely on seed funding from government, because we don't know which fund to take. So we have to take the decision. May be its not wrong to have other people, but I don't know if it is good or bad. So I need mentoring to actually make the right decision about funding"* E3P1.

Entrepreneur E8P2 at the start-up phase who was working on a GIS based property management software reported that the progress on the project was very slow, because the entrepreneur was not technical and he belonged to a real estate management business. Therefore, it was difficult to manage the technical team and monitor the progress. Entrepreneur believed that a mentor can help in managing the team.

*"I want a mentor because I am in software business but I am not technical. I know about GIS and property management therefore i am working on this project so I have a team but it's very difficult to manage the technical team when you don't know anything about it. Therefore progress is very slow and team is unable to finish project on time. I want a mentor so I can discuss with him about the project and who can help me with monitoring the progress"* E8P2.

Literature has framed mentoring relationships in terms of social problem solving (Sosik and Lee, 2002). Mentoring not only helps protégés with the business planning and market knowledge but it also provides the protégés the practical approach to solve problems that they face during the business. According to Rickard, (2007) mentoring provided wide-ranging benefits which included increasing focus, business planning, personal growth and improve problem solving skills. This helped increasing the self-confidence of the protégé. St-Jean and Audet (2012) found that mentors allow the entrepreneurs to develop the ability to develop the problem solving strategy. Entrepreneurs learn to focus on important things and targeted specific clients rather than going after any client.

### 6.1.2 Psychological Support

Psychological support as can be seen in figure 5.1 has been divided into 2 sub factors: confidence and decision making. Psychological supports of the mentor help the entrepreneurs make the right decision and boost their confidence to face the uncertainty. Entrepreneurs need to make tough decisions throughout their career. These decisions, if go wrong, can cause damage to the company. Therefore, entrepreneurs need someone who can help them in taking the right decision and boost their confidence.

Literature has documented numerous benefits of mentoring relationships. Empirical analysis found that anticipated outcome associated with psychological support drives the decision to seek mentors. Leck et al., (2009) found out psychological support was a significant predictor during her study on measuring the intention to seek a mentor. Mentoring is reported to provide psychological support to protégés. This helps mentees to grow self-confidence and develops the right mind-set to succeed (Bosi et al., 2012). Kram, (1985) proposed career, psychological and role modelling functions of mentoring that helps the protégé to acquire new knowledge and skills. Mentoring support also helps the protégé's in eliminating the uncertainty, because protégé's count on mentor's psychological support and feel more confident and safe while facing the uncertainty (Kyrgidou and Petridou, 2013).

**6.1.2.1 Confidence**

Entrepreneurs need encouragement and appreciation. They want to know whether they are going into the right direction and can be as successful as their mentors. Entrepreneur E1P1 at the conception phase reported that mentor can provide the feedback about the product that if the product is good enough to penetrate into the market. Entrepreneur believes that encouragement and feedback from the experienced mentor will boost their confidence.

> *"Because a mentor can make my work easy. Because sometimes being entrepreneur at this stage I need someone to give me some support give me moral support so that I know that where I am going to"*
> *"I want him to guide me about my product. Like if it is good and if it will be successful. It will actually encourage me if I get positive comments about my product from an experienced mentor"* E1P1.

Entrepreneur E14P1 reported about that psychological support will encourage them not to give up on their dreams, because when someone is working and wants to be entrepreneur then one needs support, so that entrepreneur do not get discouraged and do not give up.

> *"I want to be mentored because i want to be an entrepreneur and I want to know if I can be successful. I want someone to help me and give me feedback and tell me if I am taking the right steps to become an entrepreneur"* E14P1.

At the start-up phase, entrepreneurs believe the psychological support of the mentor will ensure that entrepreneur remains on the right path and do not get distracted. Mentor can set the goals, though sometimes it may get annoying to have someone who pushes you for the work, but at the end entrepreneur can become more successful and achieve more than what one could achieve without the mentor.

*As an entrepreneur, I believe that we can be successful if we have a good mentor. We can motivate ourselves. Everyday work we have to do, we don't know if we are doing it right until we get some result. So mentor will be able to tell us about the product. If it is good. I will be more motivated and will be able to achieve more than what mentor achieved.* E3P2.

At the growth phase, more entrepreneurs had prior mentoring experience. Entrepreneur E2P3 reported that mentors strengthen your philosophy. A mentor should not just guide the entrepreneur that what to do, but also tell them why. This strengthens the philosophy and perspective of the entrepreneur about the entrepreneurship.

*"I have few mentors. Even one husband and wife. They are both different. Husband is like very philosophical and it's like why this and you should have this plan, he is very high level. The wife is like, this is good, go for it, and this is good, go for it. So when I meet wife I always get different advice than when I meet husband. So I really like someone who knows how to fit us with good philosophy. So if we have our own philosophy too. They will strengthen it"* E2P3.

Entrepreneur E4P3 reported that without the mentor entrepreneurs may make mistakes, because they lack experience. Mentors' experience can make the entrepreneurial journey smoother than what it could be, if entrepreneurs have to do everything on their own.

*"I think my journey as an entrepreneur will be smoother, as there will be someone to guide me and motivate me during the course of my journey. So if I decide to do everything on my own then I may make mistakes but with a mentor I will be able to get advice on different issues that I may face"* E4P3.

Literature suggests that Psychological support helps the protégé grow confidence and self-belief (Bosi et al., 2012). Mentor offers challenging assignments which allow the protégés to develop specific competencies and build confidence in their role (Rutti et al., 2013). A research by (Donaldson et al., 2000) has shown that mentor advances a protégés career by providing emotional support and confidence, suggesting useful strategies for achieving work objectives, but literature also suggests that mentor's behaviour plays an important role in increasing the self-confidence of the protégé. Sosik and Lee, (2002) found that protégés received lower mentoring support from mentors who rated themselves higher than the protégés in terms of leadership behaviour as compared to mentors who underestimated such behaviour.

### 6.1.2.2 Decision Making

Entrepreneurs want their mentors to help them in making the right decision. It was reported that a mentor should give feedback on decisions that entrepreneur wants to take and should inform about the possible outcome of the decision. This will help entrepreneur re-evaluate the decision and change it, if needed. Entrepreneur E16P1 at the conception phase reported that a novice entrepreneur thinks too much before taking the decision. They think about the consequences that their decision may have. Therefore, mentors' advice can help the entrepreneur in taking the right decision. Entrepreneur E13P1 reported that mentor is more experienced and knows about the do's and don'ts of business, so mentor can guide if entrepreneur is going into the right direction and can guide that what an entrepreneur is supposed to do to become successful.

*"As a mentor I am expecting more moral support and encouragement. I want to know if I am doing right things and take right steps to become a successful entrepreneur. So I want my mentor to guide me that what I am supposed to do and what I am not supposed to do and something like that because they are the mentor so they should know what is right and what is wrong"* E13P1.

Entrepreneur E4P2 at start-up phase reported that mentor should help the entrepreneur in taking a decision, but he should not impose it upon the entrepreneur. It should be mutual decision where either party can argue and convince each other before taking a decision.

*"I want him to advise me that I am doing the right thing. It's not like he impose his decision on me but we can come to some consensus. May be he convince me or I convince him but I think instead of me taking decision on my own if I have someone more experienced to consult, there will be more chances that my decision get me good result. I want him to work with me as partner take mutual decisions and share the risk"* E4P2.

Another entrepreneur E5P2 reported that a wrong decision may not only cost the entrepreneur a lot, but it also means that entrepreneur loses an opportunity. Therefore it is important to consult the mentor before taking a decision even if it takes some time.

*"I need guideline to make right decision. Because decision you make costs you directly or indirectly. So its better you search and spend more time on decision making than jump and take decision. Without mentoring I might take wrong decision. This is number 1. Number 2 it will cost me a lot. Number 3 I might miss an opportunity. Like if I was expecting something will cost me 10,000 ringgit and It costs me 100,000. Then I can't really afford it. So I need to make right decision"* E5P2.

At the growth phase entrepreneur E1P3 reported that a mentor provides you a second opinion. It is because, being an entrepreneur you are your own boss, so if a situation arrives and you need to take a decision then there is no one at your level to consult with. Therefore, a mentor provides you an alternative opinion. Entrepreneur referred to his mentor who he always consult with, when facing a problem.

*"I will be able to get second opinion other than mine. Suggestion for second opinion on any problem I may face. I have a problem so who do I discuss now? And I do have a mentor, I don't want to name him I go to him I visit him and take his guide in every hurdle and in every business problem. My goal is very clear that I want to get second opinion, what advantage I am going to get the second opinion is like if he is good and we discuss our problem and sometime I follow him and sometime he is convinced ok you think what you want to do. I think it will really help me. My objective is not to get a materialistic advantage. It is more like a guideline"* E1P3.

### 6.1.3 Role Model Support

Entrepreneurs reported that they wanted to have a mentor because they want to learn from the experience of the mentor and take inspiration. Role model support is not explicit or direct support like entrepreneurial support, but it is a form of indirect support. Entrepreneurs mostly reported about a mentor who can become a role model and can inspire the entrepreneurs to become successful. Entrepreneur E2P1 at the conception phase believes that he can learn from the experience of the mentor and can get into the market by using the contacts of the mentor. Entrepreneur reported about his role model who established the company 10 years ago, entrepreneur believes that person like him can be a best mentor.

*I want someone to guide me with my business. May be someone experienced with the IT. Like my boss where I am working at the moment. He established that company 10 years ago and now he is very successful. I think person like him can be best mentor for me"* E2P1.

Entrepreneur E2P2 at the start-up phase reported that mentoring can help the entrepreneur to become successful in shorter time than what it took mentor to become successful, because entrepreneur can learn from the experience of the mentor and can avoid the mistakes that mentor made during the entrepreneurial career.

*"I think end product will be that I will be more successful entrepreneur in shorter time than your mentor. Unless he also had some mentor who guided him. He might have been successful entrepreneur without any mentor. So if learn from their mistakes and experience i can learn it in shorter time"* E2P2.

At the growth phase entrepreneur E1P3 wants to have a mentor who is successful and experienced entrepreneur in the ICT industry and who is familiar with the problems that the entrepreneur was going through. Entrepreneur believes that because being an entrepreneur one doesn't have a boss who is more senior and experienced, therefore a mentor can guide the entrepreneur towards success.

*"I have 2-3 reasons is that I have a team at the moment but I don't have similar or more experienced person than me in the team as I am the head. So I need an experienced man, I need an experienced person who has successful career, successful entrepreneur in similar business as*

*mine and who can guide me and who can tell me what he has done in the past for all these hurdles and difficulties. So it's like I need someone who is experienced and who has gone through the same situation as I am going through"* E1P3.

Literature suggests that the role model support is one of the mentoring functions (Kram, 1988; Scandura, 1992). Role modelling is a more passive function and it is seen in identification of the protégé with the mentor and also looking to the mentor as source of guidance in shaping the behaviour, attitude, values of the protégé. According to Blake-Beard, Bayne, Crosby, and Muller (2011) female protégés receive more role modelling support from their mentors as compared to male. This was found during the study to identify the influence of role models on women undergraduate students to pursue science.

## 6.2 Subjective Norms

Subjective norms reflect the individual influence on entrepreneurs' intentions towards the mentoring. This study found those entrepreneurs' intentions towards the mentoring gets influenced by peers and superiors. Though at conception phase one of the entrepreneur also reported about family influence, but it was reported only once and no respondent reported about family influence during start-up and growth phase. Therefore the family influence was not placed into a revised framework.

### 6.2.1 Peer Influence

Peer influence refers to as influence of close friends, colleagues and acquaintances on entrepreneurs' intentions towards mentoring. Entrepreneurs reported the positive mentoring experience of their friends, colleagues and partners which influenced their intentions to seek a mentor. Entrepreneur E9P1 at conception phase reported about how mentor helped one of her friend. Mentor helped the entrepreneur by connecting her with other entrepreneurs and gave the business ideas.

> *"One of my friend also have a mentor so she said that her mentor helped her in giving ideas about business and also gave her references about other entrepreneurs"* E9P1.

Another entrepreneur E17P1 reported that one of his friends asked him to seek a mentor, because that way entrepreneur could avoid mistakes.

> *"One of my friend also advised me to go for a mentor, he said if I want to avoid any mistakes,*
>
> *then I should consult someone who is experienced and ask about advice"* E17P1.

Entrepreneur E4P2 at the start-up phase reported about friends who were mentored by an entrepreneur they met during a radio interview. Mentor helped the group of friends in getting business for their event management company and also connected them with different artists.

*"I know my own friends they had a mentor. Though they are young but business they are doing is more suitable for young generation like event management artists and music industry. Mentoring really helped them in being successful in getting business. We used to work at a radio station and One day there was big personality at radio. He took few people under him including me and he taught us how to run the show how to book artist how to manage events and he has his own company and he is doing very well. When he mentored us we learned a lot from him and one of us managed to open his own event company and he is doing very well right now. He is making about 5 figures each month right now. So I have seen this method working"* E4P2.

Entrepreneur E7P2 reported that mentor of one of his friend inspired him to seek a mentor. Mentor played a major role in the success of friend and forced the friend to work on his project. Both mentor and friend of the respondent are now business partners.

*"Actually one of my entrepreneur friends advised me to have a mentor. He has a mentor and his mentor forced him on working on one of his project because my friend was very good in IT so his mentor and him both became partners in that project"* E7P2.

At the growth phase entrepreneur E1P3 reported about a senior incubate who later on became a good mentor. Entrepreneur believes that every entrepreneur has some ideal who they try to follow in order to improve their chances of being successful.

*"I have a mentor I call him guru ji, I don't want to name him. He also has a mentor and he is pretty successful. I think everyone has it. Directly or indirectly they do have someone ideal or someone they are following and when they follow they try to improve things. My mentor did this why don't I improve things more. Or maybe they avoid those mistakes which their mentor has done"* E1P3.

People often observe behaviour of peers like friends, colleagues and acquaintance and it is more likely that they adopt that behaviour if the outcome of such behaviour is valued (Engle, Schlaegel, & Delanoe, 2011). Literature suggests that friends often influence the intentions towards certain behaviour. A study of entrepreneurial intentions in 14 different countries showed that friends/peers had significant influence on entrepreneurs intentions in 6 countries (Engle et al., 2010).

### 6.2.2 Superior Influence

Superior influence is used to identify and explain the influence of referent that is on superior position than the one who is being referred. Superior influence on entrepreneurs' intentions towards the mentoring was mainly exerted by the teachers. This was because being an entrepreneur, respondents had no supervisor or manager e.g. nascent entrepreneurs who were also M.Sc. IT Entrepreneurship students at Universiti Teknologi Malaysia were influenced by their lecturer. Entrepreneurs E11P1, E13P1 and E14P1 reported that lecturer at the university asked the students to seek a mentor, because university cannot help each and every students with their products. University can equip the students with generic skill, while a mentor who is in the same business can provide the required skill set and support to the entrepreneurs.

*"(Lecturer Name) advised me to have a mentor. He said that in university we cannot help everyone with their project. We can just provide general support. So if you want more information and technical support you should find a mentor who can help you"* E11P1.

Entrepreneur E4P2 reported about the mentoring experience at the work, before entrepreneur started the business. Entrepreneur reported that his previous employer and mentor also became the investor in the company.

*"I know my own friends they had a mentor. Though they are young but business they are doing is more suitable for young generation like event management artists and music industry. Mentoring really helped them in being successful in getting business. We used to work at a radio station and One day there was big personality at radio. He took few people under him including me and he taught us how to run the show how to book artist how to manage events and he has his own company and he is doing very well. When he mentored us we learned a lot from him and one of us managed to open his own event company and he is doing very well right now. He is making about 5 figures each month right now. So I have seen this method working"* E4P2.

Entrepreneur E2P3 at the growth phase reported about the manager and at work before the entrepreneur started the business and that how she changed the philosophy of the entrepreneur.

*"When I did my bachelor degree I didn't have any mentor I did it alone except one of the computer shop guy. At that time I used to sale computer and get 250 margin. So the computer guy told me why you just sale computer and get 250. Come to us we will train you and you can not only sale computer but also assemble and repair. You can earn more than and you can manage your own team. We will give you supply. He also became a very good mentor for me. I would like a person who notices me. Yeah this guy is ok why not I become his mentor. So that times Mr Yap and his younger brother. They saw me and they actually offered me coaching and mentored me in sales and assembly. In return i became their supplier. So I'm very lucky that time"* E2P3.

Literature analysis also suggests that superior influence is mainly exerted by teachers and role models. As the principle actor and being perceived as the formal authority, instructor/teacher can influence the student's intentions towards mentoring. Lecturers' proactive or passive involvement can encourage or discourage students' intentions to perform a certain behaviour (Yang, Li, Tan, and Teo, 2007). A role model of individual is also being acknowledged as influential factor in explaining the reason for the choice of occupation and the career. (Bosma, Hessels, Schutjens, Praag, & Verheul, 2012) e.g. if an individuals' role model is someone like Steve jobs or Bill gates then it is highly likely that these role models also influence the individuals' intentions towards entrepreneurship.

### 6.3 Environment

Environment construct has been adopted from institutional theory (DiMaggio and Powell, 1983; W. Scott, 1987). Institutional forces, normative isomorphism, coercive isomorphism and mimetic isomorphism often play an important in shaping the intentions towards certain behaviour (Leaptrott, 2005; Scott, 1995).

Environmental influence was not part of the initial theoretical framework. This construct was adapted from institutional theory. Analysing the contents, it was observed that entrepreneurs reported about organizational influence on their intentions towards the mentoring, along with the individual influence. Literature suggests that environmental influence includes the normative influence, coercive influence and mimetic influence. The study found out that normative and coercive influence was reported by entrepreneurs. Entrepreneurs reported that their intentions were about mentoring were influenced through educational institutes, online media (Normative Influence) and technology incubators (Coercive Influence). Mimetic influence which is the pressure/influence from competitors to seek mentoring was not reported during any phase.

### 6.3.1 Normative Influence

Normative influence refers to as influence on entrepreneurs' intentions to seek a mentor, exerted by educational institutes and media (print, television, and internet). Nascent entrepreneurs at the conception phase reported that they wanted to seek a mentor, because university wants every student to have an industry mentor who can help them with their products. Entrepreneur E10P1 reported that university has provided the entrepreneurs, the list of mentors. Everyone can select either from the list provided by the university or can seeks a mentor and register it with the university.

*"In the class my teacher told me about it. Actually one of my friends asked in the class that how she can get mentor, what if she gets a mentor how she can assign it. I mean university have mentoring program so university want everyone to get a mentor or select a mentor from university, so she asked how she can get it. (Lecturer Name) said just give his or her name and position to register it with university and university will officially assign"* E10P1.

Entrepreneur E14P1 reported that about the influence online media and online mentoring forums in Malaysia which provide the mentoring services to the nascent entrepreneurs. Normative influence may not always yield good results. Entrepreneur E8P1 reported that he was assigned a mentor during high school, but it didn't work out because of the compatibility issues. At the start-up phase entrepreneur E2P2 reported about that his intentions were influenced during the entrepreneurship course at the university. Students were told about the benefits of having a mentor and that how important was the mentoring for motivation and securing the funding.

*"I studied entrepreneurship at the university. There we were told about mentor. They told us how this relationship works. It's more like a friendly relationship. I learned that mentor should be someone who motivates you and get you funding and stuff like that"* E2P2.

Another entrepreneur E3P2 reported about the online media and mentoring forum in Malaysia called TEAM which helps the struggling technopreneurs. At the growth phase entrepreneur E5P3 reported about entrepreneurial development organisation MaGIC (Malaysia Global Innovation and

Creativity Center) Malaysia, which advised the entrepreneurs to seek a mentor who can help the entrepreneurs with the products.

*"I have always been reminded about the benefits of having a mentor during my journey as an entrepreneur, like at the launch of MAGIC Malaysia in Cyberjaya, (Name) advised everyone that if they wanted their start-ups to succeed then they should find a mentor"* E5P3.

Literature suggests that the normative influence may originate from within or outside the organization. Social networks comprising of both members of organizations and non-members can be source of normative influence and also source of information. Normative influence can also come from industry and professional organizations that regularly seek voluntary compliance with standards for smooth operations (Leaptrott, 2005; Scott, 1995). External entities such as government, higher education institutes, Media can exert implicit normative influence on entrepreneur's intentions towards mentoring.

### 6.3.2 Coercive Influence

Coercive influence is exerted by the government bodies for the compliance with the standards. Entrepreneurs who reported about the coercive influence were incubates in different technology incubators on Malaysia. At the conception phase entrepreneur E4P1 reported that their team at incubator MagicX were assigned a mentor. Mentor provides the technical support on product development. Entrepreneurs were seeking a mentor who could provide them business and marketing support.

*"I and my partners already have a mentor provided by MagicX, (Name of Mentor). He shares a lot of his experience with us, but that's about the product development. It's more like a technical mentoring"* E4P1.

Entrepreneur E3P1 reported about the mentoring and coaching session arranged by the incubator, which influenced the entrepreneur's intentions towards mentoring. Entrepreneur E1P2 at the start-up phase reported about the seminar arranged by the incubator, where a Chinese entrepreneur talked about the mentoring and its importance. Entrepreneur reported that he wants to seek a mentor from outside of the incubator and don't want join the mentoring program offered by the incubator, because mentoring program in the incubator is not effective.

*"I attended a class. It was more like a seminar where they told us about importance of mentoring. So that's where I felt the need of mentoring.* **Who was the speaker and what did he say about mentoring.** *It was seminar here arranged by incubator when I joined here. I attended because speaker was this Chinese entrepreneur I know. I liked the way he explained and he told us how mentoring can help us in marketing and getting more business and also told us about mentoring program of this incubator.* **So you didn't join that mentoring of incubator.** *Not really I didn't contact. I think many didn't. Because you know in incubators it's just formality. They just tell these fancy mentor things so they can tell this to govt that they have a program. Beside that I don't know I never joined but I do feel that I need a mentor but not the one in incubator"* E1P2.

Entrepreneur E5P2 reported that mentors provided by the incubator are not entrepreneurs, but professionals who mentor the entrepreneurs, because it is their job. At the growth phase entrepreneur E1P3 reported about the past experience with the mentor, provided by the incubator. Entrepreneur reported that relationship did not work out because it was very formal and mentor was more like an instructor who gives same instructions to every incubatee.

*"When I started business mentor was provided by incubator but it didn't work because it was kind of formal relationship. It's been long time. It was because he had to look after everyone in incubator. It was more like some instructor who tell how to do things. You know like same instructions to everyone in incubator. It doesn't work that way. He should have addressed problems differently and individually not like in group. Mentoring is not 9-5 job"*E1P3.

Literature suggests that the coercive influence is exerted by regulatory forces for compliance with the standards (Gholami, Sulaiman, Ramayah, and Molla, 2013). Under coercive influence the individuals make rational choices to further then own best interest by maximising rewards and minimising any action by regulatory forces, due to noncompliance. One advantage for organizations or individuals, subject to coercive influence is the legitimacy and recognition of existence and compliance by regulatory forces (Hirsch, 1997). Requiring the entrepreneurs to participate in the mentoring program could affect its success. Voluntary participation in the program would be seen as positive program attribute (Horvath et al., 2008).

## 6.4 Surface Level Characteristics

Surface level characteristics overt biological characteristics that are typically reflected in physical features and are immediately observable (Tyran and Gibson, 2008). Research by (Ibarra, 1995) on interpersonal networks suggest that individuals may not be inclined to form friendship and network with people of different race and gender. Bates (2003) suggests that gender and race diversity will have a negative effect on identification in a mentoring relationship. Surface level characteristics refer to readily detectable attributes of mentor. (Rutti et al., 2013) also included sexual orientation and disability as surface level characteristics. The researcher identified five different surface level characteristics which influenced the entrepreneur's intentions towards the mentoring. Age, Gender, Race, Language and geographical distance were identified as the surface level characteristics which entrepreneurs were looking for in a mentor. Table 4.6 shows the frequency of different surface level characteristics of mentor. according to (Rutti et al., 2013; Turban et al., 2002) surface level characteristics matter at the start of the relationship, but as the relationship develops, protégé ignore the surface level dissimilarities.

Surface-Level Characteristics significantly influence the entrepreneurs' intentions towards mentoring during the conception phase. It was explored that age was the most important factor. Entrepreneurs wanted someone who is young and a successful entrepreneur. They thought that being young means that mentors will be more aware of latest technologies and trends in the ICT sector. Age will also bridge the communication gap between mentor and protégé, because due to small age gap both will be friendly. Male entrepreneurs wanted male mentor while female entrepreneurs were not really concerned about the gender. It was may be because there are more successful male IT entrepreneurs as compared to females. Language and race were equally important. Entrepreneurs were more into same race mentors. Some entrepreneurs believed that cross race mentors will be more helpful as they will bring more knowledge, but overall entrepreneurs wanted mentors of similar ethnic background. Entrepreneurs wanted mentors to be based within certain geographical distance so that both mentors and protégé could meet face to face. Though they also wanted to tools for communication like Skype, Facebook and messaging services to get into touch with a mentor, but they thought that face to face meeting once in a while is important.

At the start-up stage Surface level characteristics of the mentor were not as significant as they were during the conception stage. Entrepreneurs who reported about these characteristics did not report that these characteristics as mandatory. This indicates that surface level characteristics of mentor are more important for nascent entrepreneurs than start-up entrepreneurs. Entrepreneurs thought that age is important because it determine the deep level understanding between mentor and protégé. Gender was not significant as only one entrepreneur reported about it, that too about its insignificance. The race was important because having the same culture and ethnic background means greater liking for each other. Entrepreneurs thought that mentors and protégé should be fluent in the language they communicate. While, entrepreneurs' preferred minimal geographical distance for frequent face to face meetings

Influence of Surface Level Characteristics on entrepreneurs' intentions towards the mentoring was not as significant as it was during the conception and start-up phase. Race and language were not reported during this phase. The race was reported only once, but from a mentor's perspective where entrepreneur talked about the race problem of mentor instead of himself. Though, qualitative data cannot be quantified, but looking at the pattern one can say that age, gender and geographical distance still has influence on entrepreneurs' intentions at growth phase.

### 6.4.1 Age

Age of the mentor influences the entrepreneurs' intentions to seek or not to seek a mentor. It was explored that technology entrepreneurs want a mentor who is young and successful, because entrepreneurs believe that an older mentor may not be away of the current trends in the ICT industry. Entrepreneur E12P1 at the conception phase reported that mentor should be older, but not too old because it will be difficult to understand each other.

*"He should be experienced but I don't want too old but yes older than me. Old people are different soul. It will be difficult to understand each other"* E12P1.

Entrepreneur E3P1 reported that older generation do things in traditional way and not like the young generation who are more into gadgets and smart phones.

*"I will prefer someone who is older. Between 30-40 because older than 40 you know we are generation y. older generation tends to think the traditional way of doing things. New*

*generation is more about gadgets, animation, games. For me I prefer that range of age for mentors. Also because mobile application is new thing so I think people above 40 will have less exposure about it*"E3P1.

At the start-up phase, entrepreneurs' had different perceptions about the age of the mentor. Entrepreneur E4P2 reported that age doesn't really matter as long as both mentor and protégé get along well. While entrepreneur E5P2 reported that one needs the right person, either young or old, because young mentor may be more careless and don't give much time, while older mentor may be more sensitive to how protégé should behave. Therefore one needs the right mentor, whether young or old.

*"Age, I don't really know. It's always question for me. May be if he is too old he is more experienced. He explains more to me but also when people get old they become more sensitive and there might be a problem. If he is too young he may be more careless and will always be like. Never mind, never mind, and this never mind and carelessness may be cost me big time. So yes I think age does matter because experienced guys work for their own self and I know young guys are much more practical and they get things done without much problem, easier than others. So I need to find the right person. Need to test the person first if he is really true or not"* E5P2.

At the growth phase entrepreneur E3P3 reported that mentor shouldn't be older than 50 year old, because older people may not be aware of current trends, while young mentor may have attitude problem.

*"I think age is very important he shouldn't be older than 50. Because If he is too old than most likely he is not familiar with current trends. Compare to young they also will be more attitude problem but it depends on business other business don't care about age but IT is age sensitive"* E3P3.

Though protégés want similarity in other surface level characteristics like Gender, Race, Language and Distance, but they want dissimilarity in age. It is because protégés think that a mentor of similar age as them may not be a successful entrepreneur. Entrepreneurs also don't want a mentor who is too old because they feel that an entrepreneur who is too old might not be tech savvy and may not be familiar with latest technologies and trends.

Literature analysis suggests that the concept of a mentor has been associated with seniority and experience (Baldwin, 1998), but Finkelstein, Allen, and Rhoton (2003) maintained that age difference between mentor and protégé should be between 8 to 15 Years, as protégé may not feel comfortable with an older mentor due to age difference and a friendly relationship may not develop as this relationship will take on qualities of parent and child relationship.

### 6.4.2 Gender

Entrepreneurs want to have mentor of same gender, because they believe that it will be easier to communicate and share things. More entrepreneurs at the conception phase reported about gender of the mentor being a factor which affects their intentions towards mentoring. Entrepreneur E10P1 and E11P1 at the conception phase wants to have mentor of same gender. Entrepreneurs believe that mentor of same gender shares same thinking and can be more friendly than mentor of opposite gender.

*"I am more into male because we can talk something and we can share same thinking. May be we can be friendlier when talking"* E10P1.

Entrepreneurs E3P1 and E6P1 reported that although gender doesn't matter, but they would prefer mentor of same gender. Entrepreneurs at the start-up phase didn't report about the gender. At the growth phase 2 entrepreneurs reported about the influence of gender on their intentions towards mentoring. Entrepreneur E2P3 wants to have mentor of same gender, because of his past experience. Entrepreneur believes that in case of opposite gender there will be some limitation.

*"I prefer male mentor because I am that kind of person who always call anytime. So I think gender is important.* **You said gender is very important what if someone have experience but she is female?***. I don't mind in that case but as I said mentor-mentee is heart to heart relation. It's not like you my mentor and I am your mentee. So if it's a female there may be some limitations. So yes I will still prefer experience even if it's a lady but maybe I have to modify the way I communicate. May be we fix some time and mode of communication"* E2P3.

Entrepreneur E3P3 reported that business is gender neutral, therefore it will be good to have a mentor of opposite gender as it will give the perspective of opposite gender.

*"Gender I think it will be good if I have a female mentor but I don't really mind. Female mentor can give me perspective of opposite gender that how they think because business can't be gender bias it is gender neutral"* E3P3

Literature suggests that the similarity between the mentor and the protégé´, either real or supposed, could affect the quality of the mentoring relationship. Research shows that people form first impression of others based on the appearance of the people, like gender and race, and classify the people based on their gender and race similarity (Brickson, 2000). Much has been written about cross gender and cross ethnic dyad of mentor and protégé (Kram, 1983; Ragins and Cotton, 1991; Ragins, 1997; Rutti et al., 2013; Thomas, 1990; Turban et al., 2002). Literature suggests that gender, race dissimilarity vary during the course of the relationship and as relationship develops, deep level characteristics like attitude, values, and skills take over the surface level characteristics.

### 6.4.3 Race

Influence of the race or ethnicity of the mentor also affects the entrepreneurs' intentions towards mentoring. It was observed that entrepreneur want mentor of similar race. Some entrepreneurs also reported to have wanted the mentor of a certain race, because they are more entrepreneurial. Entrepreneurs at the conception phase were more concerned about the race of the mentor than entrepreneurs at the start-up and growth phase. Entrepreneur E8P1 reported that mentor should be of a different race, because mentors of certain race have more knowledge.

Entrepreneur E10P1 wants to have mentor of Malay or Chinese race, because entrepreneur believes that Chinese people are more entrepreneurial.

*"He should Malay or Chinese. Well Malay because we are same races same so we understand each other's values. Chinese because you know in Malaysia they are business people and majority of the entrepreneurs are Chinese"* E10P1.

Entrepreneur E1P1 and E17P1 wants to have mentor of same race. Entrepreneur E5P2 at the start-up phase who is a Middle Eastern living in Malaysia reported that it is difficult to communicate if person belongs to different race, because different race also mean different culture. Therefore it is important that mentor and protégé understand each other.

*"I want someone who is accessible face to face and not someone who live far away. Because then there will be a problem. We might don't understand each other. Especially if we have a cultural problem and language problem than long distance mentoring will add more problems into it. So if it's face to face than it will be better. You like in Malaysia there is a communication problem. Every one's way of communication is different. If an African talk to Chinese. Chinese will be talk why are you shouting. This is the way they talk he isn't shouting. So if mentoring is on the phone or internet and mentor is of some other ethnicity and culture there may be a problem. He might get me wrong when I don't really mean it"* E5P2.

At the growth phase entrepreneur didn't report about his point of view, but of mentor's. Entrepreneur E3P3 reported that a mentor should not have a race problem, because otherwise mentor may not be sincere with the entrepreneur.

*"I think mentor should not have a race problem. Like in Malaysia. If my mentor have race problem he will not be sincere. He might act like he is helping me but he is not"* E3P3.

Literature analysis suggests that though much has been written on importance of diversified mentoring relationships, there is little empirical work on such relationships. Empirical research on matching of mentor and protégé has been limited (Blake-Beard et al., 2011; Hu, Thomas, and Lance, 2008). Studies about racial matching show no more consistency than studies on gender matching but few studies suggest that protégé receive more support from mentor of same race. A study by (Ortiz-Walters and Gilson, 2005) on doctoral students found that, students received more psychosocial and instrumental support from mentors of same race than from different race. They also felt more comfortable and satisfied being protégé of mentor belonging to same race. Similarly Latino students who were attached with Latino mentors were found to be more satisfied with their mentors and found them cooperative in their personal and career development than students paired with mentors of different race (Santos and Reigadas, 2002). Studies of informal mentoring relationships show that when protégé choose their own mentors, they incline to pursue mentors from the same ethnic or cultural background (Sánchez, Colón, & Esparza, 2005).

### 6.4.4 Language

Language used between mentor and protégé for communication plays an important role in influencing the intentions of entrepreneur towards mentoring, because it doesn't matter how skilled a mentor is, if one cannot communicate in the language understood by the protégé of vice versa then relationship won't take off. Entrepreneurs want to communicate with their mentors in mother tongue. Entrepreneurs also doesn't mind communicating in English, but it was observed that entrepreneurs in Malaysia were not that fluent in it. Entrepreneur E3P1, E5P1 and E7P1 at conception phase reported that they would prefer to communicate in their mother tongue, though they don't mind communicating in English, but they felt that it will be difficult to convey the right message in language other than their native language.

*"I prefer someone who speaks Bahasa Malay. English is also okay. I feel more comfortable when talking in my mother tongue"* E5P1.
*"Language does matter because I feel comfortable when speaking in my mother tongue. I feel easy that way"* E3P1.

At the start-up phase only two entrepreneurs reported about the importance of language. Entrepreneur E5P2 and E7P2 reported that language is important, therefore mentor should be native. It was observed that Chinese entrepreneurs had an advantage, because they knew Chinese and as well as Bahasa Malay and English. On the other hand Malay entrepreneurs only spoke Bahasa Malay and English.

*"Language does matter sometimes but I think if we can understand each other than it's not a problem"* E5P2.
*"I think language can be the problem. Mentor should be native and should speak Bahasa Malay or Chinese. Using language that entrepreneur familiar may get a better relationship"* E7P2.

At the growth phase, language of the mentor was not reported as a factor that influences the entrepreneurs' intentions. Literature analysis suggests that Language barriers can have significant impact on effective communication between mentor and protégé. Both mentor and protégé must understand of what is being written and/or said (Feldman, 1999). In a multicultural society, language is usually considered as an expected barrier in communication. During a mentoring session both the mentor and protégé should have Command over the common language used during the interaction. Both should settle on the language to be used all through mentoring session or period. Athey, Avery and Zemsky (2000) found that mentoring can be more effective if mentor and protégé share the same culture and language.

### 6.4.5 Geographical Distance.

Geographical distance between the mentor and protégé play an important role in influencing the entrepreneurs' intentions towards mentoring. Entrepreneurs want a mentor who lives in close proximity, so that a face to face meeting is possible. Few entrepreneurs believe that with the emergence of social media and conferencing tools it is not important for the mentor to be based within a certain geographical distance, because protégé can still maintain contact with the mentor over internet. Entrepreneurs E8P1 E10P1 and E13P1 at the conception phase reported that, although one can communicate using tools like Skype, WhatsApp and Facebook, but it is important to have a face to face meeting once in a while. Therefore mentor should be based within an accessible distance.

*"I will like face to face but whatsapp and message is also ok but more face to face because in message we cannot express so in face to face we will express more"* E10P1.

*"We can communicate on Skype and phone so as long as he is giving me time when I need I think anyway of communication should be fine but he should live in Johor because if sometimes I want to meet face to face then we can meet"* E13P1.

At the start-up phase entrepreneur E5P wants to have a mentor who live within accessible geographical distance. Entrepreneurs believe that face to face meeting is very important, because mentor and protégé can understand each other better from the expressions that what other person wants to say. Entrepreneur E8P2 believes that mentor and protégé should have a face to face meetings at the start of the relationship and once both understand each other, then communication can happen over internet or phone.

*"He should be based within Johor so we can discuss about it. Or maybe if he is living in some other city then he should first see and understand the project so that when we discuss about it on phone or internets he know that what I am talking about and he understand it"* E8P2

Entrepreneur E6P2 reported that the mentor should be living in some other country because, that way he can help in getting projects from other country or shall be living in a different state within the country.

*I would like it if my mentor Is someone in other country. May be he can get me projects from there. Like outsourcing. Even in Malaysia if I can get someone from other states and east Malaysia then it will be good. May be market over there is good and I can get clients.* E6P2.

At the growth phase entrepreneurs E3P3 emphasized that mentor should be within the same city, because nature of their business required someone who understands local market. Entrepreneur E4P3 also believe that face to face meeting is more effective, therefore mentor should be based within the same city.

*I will prefer someone from Kuala Lumpur or Cyberjaya, because nature of my business is different. My clients are local and I provide them support by going there. So I need someone who is not that far from here and know about the clients in the city"* E3P3.

Literature analysis suggests that the geographical distance between mentor and protégé determine the frequency of interaction between the two (Tammy D. Allen, 2010). Empirical evidence also suggests that physical/geographical distance between mentor and protégé plays an important role in scheduling the meetings and face to face interactions. According to Eby and Lockwood (2005) mentors and protégé both reported that geographical distance between them is one of the problem in scheduling meetings. Another study of 30 different mentoring programs examined that over half of the programs allowed both face to face and distance relationships while 40% programs were focused on only face to face relationships. It was explored that even in distance relationship programs there were at least one face to face meeting between mentor and protégé in order to facilitate relationship building (Eddy et al., 2005).

## 6.5　Deep Level Characteristics

While, surface level characteristics are immediately observable and measureable, deep level characteristics referred to as underlying psychological characteristics. These characteristics are not immediately observable, but information about these characteristics is communicated through verbal and non-verbal behaviour patterns and is only learnt through extended interactions and information gatherings (Guillaume, Brodbeck, and Riketta, 2012).

An inappropriate match with the dyad due to different values, interests and work style can have negative impact on mentor and protégé relationships (Hamlin and Sage, 2011). Therefore Mentors and protégés should share a common ground based on mutual interest and values. Before entering into formal relationship, the mentor and the protégé should become familiar and informally explain their common interests, shared values, future goals and vision. The relationship may start well if both that mentor and protégé give a high priority and take time to become familiar with one another's interests, values, and goals (Hiemstra and Brockett, 1998).

Deep Level characteristics represent the characteristics of the mentor which cannot be detected readily, but can be observed over a period of time. These characteristics include the personality, skills, experience, accessibility, partnership and religious belief. Initial theoretical framework included the attitude and trust factors, but those were replaced by personality in the revised framework because the personality of the mentor incorporates both. Religious belief was not part of the initial framework, but after analysing the data, researchers decide to replace shared values with religious values of the mentor. Researcher thinks that religious belief of a mentor plays an important role in influencing the entrepreneurs' intentions towards the mentoring initially but, as entrepreneur progresses in the career then the personality of the mentor takes over the religious belief factor. Similarly partnership factor was also made part of the final framework because some entrepreneurs reported that they wanted a mentor who can eventually become a partner. Though some of the entrepreneurs didn't agree with that and reported that they wouldn't want a mentor who want to be a partner in the business.

At the conception phase, nascent entrepreneurs wanted similarity in personality, religion, accessibility and religious belief, while dissimilarity in skills and experience. Entrepreneurs think that deep level similarity or compatibility is very important in order to create mutual understanding between mentor and protégé. This understanding will help in creating a strong bond between the two. Accessibility or availability of the mentor was the most important factor, followed by Personality, skills and experience. Religious belief was a new factor which was not part of the initial theoretical framework because empirical evidence on the influence of religion on intentions towards mentoring is not adequate. Though there are studies which talk about beliefs and values in general (Hiemstra and Brockett, 1998; Tyran and Gibson, 2008) but there is not enough empirical evidence about influence of a specific religion or belief on mentoring intentions. During conception phase, religious belief of the mentor emerged as a significant factor. Entrepreneurs reported that they wanted a mentor with similar religious background. All of the respondents who talked about religious belief of the mentor were Muslims but not all Muslim respondents reported about similar religious belief.

Deep level characteristics of the mentor at start-up stage were more significant than conception stage, except religion which was reported by only 2 of 8 (25%) entrepreneurs at start-up stage as compared to 9 of 17 (53%) entrepreneurs who reported about religion during conception phase. A new factor "Partnership" was identified during start-up stage. Entrepreneurs wanted their mentors to work as partner on their projects and share the risk and money involved in the project. Accessibility or availability of the entrepreneur was most prominent factor at this stage, because all 8 entrepreneurs reported about it during this stage. Entrepreneurs wanted their mentor to be available for advice when they needed them. Entrepreneurs also reported about the personality of the mentor as an important factor. They wanted a mentor who could help them, but not dictate them. They just wanted a mentor to advise them and let them take a decision on their own.

At the growth phase, influence of Deep Level Characteristics of the mentor remained significant. The religion of the mentor was only reported once, but entrepreneur reported about the insignificance of religion rather than significance. The experience of the mentor remained most prominent, because all of the respondents reported about it. Entrepreneurs wanted a mentor who does not dictate. Citing how mentoring relationships in the past turned sour because of the authoritative behaviour of the mentor. Entrepreneurs also talk about the skills on the mentor. They preferred a mentor who was a serial entrepreneur and strategic thinker.

### 6.5.1 Accessibility

Accessibility or availability of the mentor plays important role in shaping the intentions of an entrepreneur towards mentoring. Entrepreneurs want a mentor who is available when protégé is facing a problem or needs an advice. Entrepreneurs E1P1, E8P1 and E16P1 reported that they needed more support at the conception phase, because they were still developing their products. Therefore they want a mentor who provides more support and is readily available. Entrepreneurs don't mind communicating over the internet, if mentor is not available at certain time. Entrepreneurs E17P1 reported that mentors are usually busy with their own businesses, but they should still take time to mentor the protégé if they are entering into the relationship.

*"He should give me enough time because at the early stage of starting business I would need frequent advice from my mentor. Also at this stage I need more support so may be time becomes a problem because I want to meet with my mentor 2-3 days a week"* E1P1.
*"He should give me more time at the start because I don't have any knowledge about entrepreneurship so I want to meet often like once a week"* E8P1.

> *"He should also be more accessible. I know a successful entrepreneur will be very busy but may be once in a while we can meet and I can discuss everything but he should give concentration. Not like he gives me time and keeps busy about his own business and don't listen to me"* E17P1.

Entrepreneur E3P2 at the start-up phase wants the mentor to have a timetable. Mentor doesn't need to follow the schedule of the protégé, but should be accessible and available if protégé is facing a problem. Entrepreneur believes that being a technopreneur, one faces challenges every day and needs the support of the mentor. Entrepreneur E5P2 thinks otherwise, entrepreneur reported that mentor should fix the time according to the schedule of the protégé, because if protégé wants to complete a project within a certain time and mentor is busy then protégé may not be able to meet the deadline.

> *"There are times that my mentor is busy and I feel like I am facing problems and I can't do it alone. So when I contact him. He says I am sorry I am busy and not available. So I don't know whom I should share problem. Because being an IT-entrepreneur time is very important. Every day there is challenges so I think accessibility may be a problem. Also he should be accessible and give more time for coaching. I am not saying he has to follow my schedule, but I just want him to be there I need him"* E3P2.

Entrepreneur E2P3 at the growth phase wants a mentor who can just be in touch through different social media tools. Entrepreneur believe that mentor and protégé shouldn't just talk when protégé needs help, but they can always drop a message enquire about each other.

> *"I manage my Google calendar. Everyone can see what I am doing and where m i. but you never see that I write. Meeting with the mentor. I meet them for the project like for operational things. I want someone who can just be in touch. Technology now is advancing so unless we don't have a project to discuss about. I would just want to be in touch and say Hello and Hi"* E2P3.

Literature analysis suggests that a mentor must have time. It is important, because if a mentor cannot give proper time to a protégé when it is needed or is too busy, because they are mentoring too many protégés, then it will neither help the mentor nor the protégé. A protégé must look at the timetable of mentor and see if the mentor has enough time for them. Lavin Colky and Young (2006); St-Jean and Audet 2012) believe that the availability of mentor is not a problem as long as the meetings that do take place between mentor and protégé, are effective and efficient. Therefore, according to this research, the effectiveness of mentoring is more important than frequency. Eller, Lev, and Feurer (2014) examined what protégés wanted and they found that it was more frequent communication with their mentor. They emphasized that mentors should be available beyond office hours via email and phone. Entrepreneurs want to meet frequently during the early days of relationship to understand each other and also because at the conception phase they need frequent advice about the problem they are facing.

### 6.5.2 Personality

Personality of the mentor incorporates many traits. Entrepreneurs want a mentor who is passionate about mentoring, willing to listen and can fit one's' self into the shoes of the protégé. Entrepreneurs E10P1 and E13P1 at the conception phase reported that mentor should have willingness to mentor and should inspire the protégé, because if protégé is not inspired with the skills and knowledge of the mentor then it will have negative influence of the protégés' intentions towards mentoring.

*"May be if my mentor does not have passion anymore , maybe he have passion but maybe he doesn't inspire me anymore and I feel like he don't want to help me and he don't give me time when I need it"* E10P1.

Entrepreneur E16P1 reported that mentor shouldn't have attitude problem and he shouldn't boast about his own success, but guide the entrepreneur towards success.

*"I think its attitude. Mentor shouldn't have attitude problem he should always look at me like he was at my stage. He shouldn't talk about how successful he is but about what he did to become successful and what he avoided"* E16P1.

At the start-up phase, entrepreneur E4P2 reported that mentor should have patience, because entrepreneurs at this phase are novice and they keep asking questions which may make no sense. Therefore, mentor should be patient enough to listen to their queries like a teacher listen to students' queries in the class. Entrepreneur E5P2 reported that mentor shouldn't be like an instructor who just orders the protégé to perform different tasks. Mentor should also tell the reason that why protégé should perform certain task. Entrepreneur E2P2 reported that mentor shouldn't impose the decision. A mentor's task is to guide and give advice and let the protégé decide on own.

*"My point of view of qualities will be very different. Because I am still young. I will look for patience. He should be very patient. Because he will be teaching like you know a teacher in primary school how they teach young kids from A to Z, numbers and everything. Because that's their role, same goes for mentor. They are exposing us about business and entrepreneurship. When we are coming from world which has nothing to do with it. And he has experience where he can understand everything and can relate everything so he needs the patience to answer all of our questions even though if they are studied and even though it doesn't make any sense. They should give good answers, because his answers are going to motivate us or demotivate us"* E4P2.

*First of all it needs to be black and white. There shouldn't be any hidden things. He should be open. Second he shouldn't be like instructor. Do this, do that and bring this paper. It doesn't make sense. It has to be like I need to get this paper in order to get this. You know like he should tell us why he wants this paper. You know like explain everything so I have the knowledge like why am I doing this and think whether I should do this or do something else which I think is right.* E5P2.

At the growth phase entrepreneur E1P3 reported that a mentor should be loyal with the protégé and should not have any conflict of interest, because if mentor look at the protégé as someone to benefit ones' own business, instead of protégés' then it is highly likely that mentor will give advice that would benefit the mentor and not the protégé.

*"I am looking for someone who is kind of loyal to me or I should say loyal in giving me suggestions. So if person is greedy he would think of his benefit first while giving me any suggestions and not mine. So I want someone who don't have any conflict of interest with me. I would say he should be sincere because if I feel that my mentor is sincere in mentoring me rest of the things can be managed"* E1P3.

It has been suggested that personality of the mentor affects the involvement of protégé in the mentoring relationship (Chao, 1997; Kram, 1983). The personality features of mentor that are related to mentoring support provided by mentor have important practical implications. Ragins, Cotton, and Miller, (2000) suggested that quality of mentor plays an important role in shaping the protégé's attitude towards the behaviour, their career and their organization. It was advised that individual characteristics that relate to one's ability to become mentor need to be identified. Bozionelos (2004) identified that relevance between personality of the mentor and mentoring support provided is limited, however it is possible that personality of the mentor influences more strongly on intentions or motivation to provide mentoring support than it influences the actual mentoring support provided by the mentor.

Literature suggests that attitude and behaviour of the mentor with the protégé is also part of the personality (Morrison, 2009). Protégés like to emulate the attitude and behaviour of their mentors; therefore they feel more comfortable approaching their mentors for advice. Mentors whose protégé report higher level of liking towards them will be more likely to offer greater mentoring support than protégés whose mentor report lower level of liking (Lankau, Riordan, & Thomas, 2005). Beaunae (2009) identified that attitude of mentor as helper was greatly appreciated by protégés, which resulted higher learning interactions between the two. It was concluded that protégés appreciated the attitude of their mentors because they received support and encouragement from them. They were more inclined towards their mentors due to this positive attitude, encouragement and positive learning interactions.

Five factor model identified five different human personality traits: Neuroticism, Extraversion, openness, agreeableness and conscientiousness (Digman, 1990). Neuroticism has been characterised as anxiety, inhibition, negative mood and being self-focused. This makes the mentors who score high on neuroticism less likely to approach to entrepreneurs to provide them mentoring support.

Extraversion can be defined as being spontaneous, active and intimate in social interactions. Therefore an extravert mentor will more likely approach the potential protégé and will interact with them beyond formal work activity. Extraversion is related to interpersonal competence of mentor. Literature suggest that protégés also show preference for mentor with high level of interpersonal competence (Bozionelos, 2004).

Openness is cognitive and emotional flexibility and being open to new ideas and experiences. Therefore, mentors who score high on openness will more likely to listen to ideas of their protégés and accept their idiosyncrasies. These mentors are more tolerant to diverse views and ideas of their protégés.

Agreeableness can be defined as care and concern for others and also having trust in the protégé. Hence agreeable mentors will be more inclined to provide advice, help and support to less experience protégé. Also agreeable mentors will not feel threatened by their protégés and will be more willing to share their knowledge with others.

Conscientiousness is about sense of duty and being adherent to moral principles. Therefore conscientious mentors feel it their responsibility to provide support to less developed protégés. Conscientiousness lead to identification with the work effort of subordinates which increases the likelihood to mentor them (Allen, Poteet, and Burroughs, 1997).

### 6.5.3 Skills

Skills of the mentor refer to the degree of expertise that a mentor possesses in a certain realm. Entrepreneurs reported about different set of skills at different phases of entrepreneurial career. At the conception phase, entrepreneurs reported that they wanted their mentors to possess technical (expertise in system development, mobile application development etc.) and managerial skills. Entrepreneur E11P1 at the conception phase wants to have a mentor who is more technical. It was because nascent entrepreneur were already running a business and lacked technical skills that entrepreneur wanted to develop for another business. Entrepreneur E12P1 is developing a mobile application for interactive learning and believes that a mentor who is expert in mobile advertisement and marketing. Entrepreneur believes that these expertise will help make a better application.

*"He should be technical and should understand my requirements. I need technical and information support so I want my mentor to be more technical. I am already running a business so I know that there is a requirement so I can use this system and sale it, but I don't know how to develop it"* E11P1.

*"If I get the right mentor who really knows about this interactive learning then I think I will be able to get funding and also I will make my app better because my mentor can tell me that what is good and what I should have in my app. He should Also be expert in design and marketing, I mean advertisement of mobile apps, I think mobile app or IT marketing is different than product marketing so he should be related to IT"* E12P1.

Entrepreneur E7P2 at the start-up phase reported that mentor should possess basic organisational skills and should know about the mistakes that a typical start-up should avoid. Entrepreneur E3P2 and E6P2 reported that mentor should possess ICT skills, preferably in software development, because otherwise mentor may not understand what problems an entrepreneur is facing.

*"He should have known how of basic organization work and common mistake for a start-up. Being entrepreneur doesn't mean that I know everything about business, so he should guide me away from mistake and it is vital"* E7P2.

*"Another problem can be if I feel that he doesn't have enough knowledge or he doesn't understand me and my problem. May be he is ICT but you know ICT industry is very big. I want him to be related to software development and project management"* E6P2.

At the growth phase entrepreneur E2P3 reported that mentor should be a strategic thinker who knows about the competitors and the market, because if a mentor does not know about the existing market players then entrepreneur wouldn't get the needed support. Entrepreneur E4P3 wants a mentor who is a serial entrepreneur, because if a mentor has successfully started businesses then such mentor can bring value for the company.

*"If you ask me the qualities. I would say he should be strategic thinker. He really knows the trend of the business. He knows all the player in his domain. He knows about demand supply and market. Like competitors. This is like jobs based or I should say professional qualities"*. E2P3.

Literature analysis suggests that the skills of a mentor are one of the many deep level characteristics of mentor. A mentor should possess certain skills that are required by the protégé. The skills of a mentor depend on what the protégé wants to learn from them. A good mentor possesses outstanding communication skills and is able to use their skills to mould the personality and style of the protégé. A good mentor will always provide the protégé with challenges that will nurture professional development and a feeling of achievement in learning the field (Johnson, 2003). Vonk (1993) found that a mentor should possess interpersonal skills such as problem solving, listening, encouraging etc. Though these skills are not same as what explored in this study, but it is because there is little empirical evidence available about the skills of entrepreneurial mentor.

### 6.5.4 Experience

An experienced mentor can make the road to the success much smoother for the protégé. Entrepreneurs want a mentor who is experienced, but they also realize that an experienced mentor is hard to get. Entrepreneur E16P1 at the conception phase wants a mentor who is experienced and successful entrepreneur. Entrepreneur reported that there are many experienced mentors, but they are not as successful in their career. Therefore entrepreneur would prefer a mentor who is less experienced but successful over a mentor who is more experienced but their success does not reflect their experience. Entrepreneur E12P1 wants a mentor who is not that experienced, because being experienced means a mentor must be old and entrepreneur wants a mentor who is young.

*"I will prefer someone who is a very experienced entrepreneur. I mean experienced in a sense that he is very successful and his experience should reflect his success. There are many entrepreneurs who are very experienced but are not earning as much"* E16P1.

*"He should be experienced but I don't want too old but yes older than me. Old people are different soul. It will be difficult to understand each other"* E12P1.

Entrepreneurs E1P2 at the start-up phase wants a mentor who is not only experienced, but his experience should be in the ICT industry. Entrepreneur E5P2 reported about a mentor who is experienced and is also in touch with the current trends and know about the market.

*"He should be experienced in same field. It is very important. More experience more knowledgeable in the field. So I think that's all I am looking"* E1P2.

*"He need to be experienced, he know latest updates. He also needs to be realistic and openness I say. There has to be. Otherwise you just go blind and in the end you come to know something surprising which you didn't know earlier"* E5P2.

At the growth phase entrepreneurs E2P3 and E4P3 reported that mentors' experience should be relevant to the protégés' industry, otherwise mentor may not be the right person to seek guidance. Entrepreneurs were looking for a mentor with relevant industry experience.

*"Other one as I mentioned mentor should be very positive. He should be very like he knows what to advice. Like some mentor they can advise you but I would like to know mentor who have gone through similar experience and it's not just advice. Some mentors advice you based on the cases but if the mentor have gone through process. Like If I advise you Mr. Jamshed your should do A and B and why I am saying it because I also went through it. This is the best for you. I don't mentor who comes to me and like you do this. I am like have you tried before. He is like no I haven't tried but you can try. It will work. Mentor should tell show try and do. He shouldn't experiment on mentee"* E2P3.

*"If I am working on F&B business, a mentor may have successful startup in logistic industry. Her experience may be irrelevant, and may not be able to advise me on specific issues in F&B industry. So he should have relevant experience in ICT industry"* E4P3.

Experienced mentors, as opposed to inexperienced ones, tend to be more difficult to get because they are often pursued by many aspiring entrepreneurs. The mentor's choice in selecting what he thinks would be the best protégé is important. In order to select a protégé, a mentor ensures that a protégé wants to learn to become an entrepreneur and is willing to work hard to become one (Rose, 2003). According to Allen (2003) experienced mentor is more helpful and empathically towards protégé than a non-experienced mentor. In an earlier study, (Allen et al., 1997) interviewed experienced mentors to understand the reasons they wanted to mentor others. It was examined that mentors wanted to mentor others due to two reasons, self-focused and other-focused. Other-focused motives include the desire to help others, Desire to pass along information to others and desire to build a competent workforce. While self-focused motives include the desire to increase the personal learning and feel gratification (Allen, 2003).

### 6.5.5 Religion

Religious belief of the mentor plays an important role in influencing entrepreneurs' intentions towards mentoring. Entrepreneurs at the conception phase reported that similarity in the religious beliefs would be preferred, because entrepreneurs feel easy to communicate with mentors who share the same faith as theirs. Entrepreneur E13P1 at the conception phase wants a Muslim mentor. Entrepreneur believes that every religion has different teachings and morals. Therefore, if protégé wants moral support from a mentor who has different religious beliefs then it will create problems. Entrepreneur E8P1 also wants a Muslim mentor. Entrepreneur reported that mentor being from different race doesn't matter, but mentor should share the same religious belief as of protégés'.

*"I prefer a Muslim mentor because as race everybody is same but in religion there are different teachings in Islam and different teachings in other religions so when you are Muslim and you follow someone with different teachings then you are going to have a problem with your thinking. As I said I need mentor to encourage me to give me moral support and there are different morals when it comes to religion"* E13P1.

*"I think if mentor is from some other race then it will be better. Because sometimes some race have more knowledge, so they can give me that but I prefer a Muslim mentor because I will feel easy to communicate and share things. Otherwise there may be problems which I don't share with my mentor because he doesn't understand"* E8P1.

Entrepreneur E2P2 at the start-up phase wants a mentor whose business is shariah compliant, because entrepreneur believes that being shariah compliant means mentor follows certain business ethics.

*"Mentor should be shariah complaint. Shariah Compliant means his business should not be against shariah. Like halal business.* **So he should be a Muslim?** *Not necessarily Muslim. He can be a non-muslim. You know may be Muslim. Like Muslim in heart. Shariah compliance is more about business ethics. He should keep his promises and deal honestly"* E2P2.

Religious belief of the mentor was not reported at the growth phase. It was explored that all the entrepreneurs who reported the religious belief of the mentor were Muslims. It may be because Malaysia is a Muslim majority country and most of the respondents were also Muslims. It is also possible that respondents of other faiths did not want to share about their religious beliefs.

Literature analysis suggests that heterogeneity between mentor and protégé in belief and values can affect the intentions of entrepreneurs towards mentoring. According to Tyran and Gibson, (2008) similarity in belief is more important than ethnic similarity. Though unlike ethnic compatibility which is surface level attribute and easily detectable, belief compatibility is deep level attribute and it is not immediately detectable but it takes time to unfold. Therefore before entering into a formal relationship, the mentor and the protégé should become familiar and informally explain their common interests, shared values, and future goals and vision. The relationship may have a good start if mentor and protégé give a high priority and take time to become familiar with one another's interests, values, and goals (Hiemstra and Brockett, 1998).

### 6.5.6 Partnership

Partnership between the mentor and protégé was reported only during start-up and growth phase. It may be becauseentrepreneurs at the conception phase had not started their business yet. Entrepreneurs want a mentor who can also work as a business partner and share the risk. Entrepreneur E2P2 reported that it is better that mentor and protégé also enter into partner relationship along with being mentor and protégé. This will help avoid any conflict, in case mentor wants a share in profit that entrepreneur made due to the advice of the mentor. Entrepreneur believes that mentoring is a voluntary job and mentor shouldn't ask for monetary benefits from the protégé unless mentor also shares the risk of loss which may incur by following the advice of the mentor.

*"There may be a problem of monetary benefits. Example if my company benefits from his mentoring. Mentor might want some share of profit. I think this should be some pre understanding. These things shouldn't come later. I don't mind about sharing profit but I think it depends on situation. If mentor also shared the risk like he also injected money into the project than yes he should get it but if I took all the risk than its up to me to share it or not"* E2P2.

Entrepreneur E8P2 reported about his past experience with the mentor who also worked as his partner. He reported that his mentor/partner cheated him and stole the code of software they were working on. Therefore he wanted a mentor just for entrepreneurial support and didn't want to share the project.

*"Actually I had a mentor in the past. He was my partner as well and more technical but then there were a problem between us and we separated and he cheated me and took the software. Therefore now I just want a mentor with who I can discuss it sometime without involving him directly into the project"* E8P2.

Entrepreneur E2P3 at the growth phase wants a mentor who don't want monetary benefit in return. Entrepreneur believes that mentoring is about helping the protégé and in future protégé will mentor some other protégé. Entrepreneur E3P3 reported that mentor can get the equity in the company, but one should see what value mentor brings for the company.

*"You know one of the things when our knowledge grows then we share it right? So I am successful by his mentorship then I believe that he will have a follower and if I am somewhere in life and sometime I can also help him too. Sorry but my idea of mentorship or getting guideline is more into ethical than monetary benefits. It's like he is helping me I will help someone else"* E2P3.

*"There may be a problem over certain financial benefits in return on mentoring but I think if he asks for equity in company then it is ok. I will see if he brings some value to the company then it is possible"* E3P3.

Literature suggests that there is not enough empirical evidence about mentor as a business partner in entrepreneurial research. However (Glenn, 2006) talked about partnership between mentor and protégé in educational institutes. Paper talked about how mentor protégé relationship between two teachers turned into partnership where a mentor teacher helped her protégé in with the courses and in return she (mentor) used her protégé's poems and taught it to her students. In a research study in south Africa Watson (2009) found that a mentoring service provider named Business Partner Mentors provided mentoring services to SMEs in return of partnership in the business.

## 6.6 Past Experience

Past mentoring experience has been reported during all the phases. Entrepreneurs reported that mentoring experience during academic and professional life had significant influence on their intentions to seek a mentor again. During the early phase of entrepreneurship entrepreneurs, mostly reported about their mentoring experience during early and higher education. During start-up and growth phase some entrepreneurs reported about their experience with the mentor who helped them with the business. Past experience was placed as a background factor. Figure 5.1 shows the revised theoretical framework of mentor-protégé matchmaking.

During the conception phase not many entrepreneurs had past mentoring experience during conception phase. Those who reported about past experience, reported about academic mentoring. It may be because entrepreneurs were still at conception phase and had not developed their products. Therefore, they did not have an entrepreneurial mentor. 5 of 17 entrepreneurs reported about past experience during this phase. Mentoring was mostly related to psychological support that they got from teachers in school and higher education institutes. Entrepreneur E1P1 reported about a mentor he met while in the university. It was a short relationship and entrepreneur just learned about business skills from the mentor. Entrepreneur E2P1 reported about academic mentor that was assigned by the institution. Mentor advised the entrepreneur who was a student that time in selecting the subjects and gave career advice. Entrepreneur E7P1 reported about the mentor that has been provided by the incubator. Mentor helps the entrepreneur in product development.

*"I learned about mentor in university. Everyone had mentor but it was more academic mentor and helped students with the studies. I actually liked it because mentoring wasn't about help with the subjects like teachers do. It was more kind of advice about my academic career. Like subjects I should choose and how I can improve my grades"* E2P1.

*"I and my partners already have a mentor provided by MagicX, (Name). He shares a lot of his experience with us but that's about the product development. It's more like technical mentoring"* E7P1.

Past experience with a mentor was reported by four of eight entrepreneurs during start-up phase. Entrepreneurs described about their mentoring experience when they were at conception phase, and that how they were helped by the mentor. Due to past experience, entrepreneurs at this phase were more cautious about the new mentor because 2 of the 4 entrepreneurs reported about negative experiences with the mentor. Though they acknowledged the benefits of mentoring, but they were more careful while selecting a new mentor. Entrepreneur E6P2 reported about the academic mentor who influenced the protégé's intentions towards entrepreneurship. Mentor advised the entrepreneur to start own business, because of the programming and analytical skills that protégé possessed. Entrepreneur believes that it was because of the motivational support that the academic mentor provided, due to which he could join the incubator and start a business.

*"Well I had a mentor before. He wasn't entrepreneur he was just my lecturer in the university. He convinced me that I should start a business because I was very good at programming and development. I learned some new tools which were in demand in the industry. It was because of his motivation that I joined this incubator and started business. Otherwise may be I would have just worked in some company"* E6P2.

Past mentoring experience was reported by four of five entrepreneurs during this phase. Entrepreneurs were aware of the importance of mentoring because of the role they had played in their success. Like start-up phase, entrepreneurs at growth phase also described their negative mentoring experience. An Entrepreneur reported that his mentor wanted to impose his decisions upon him, which a mentor shouldn't do. A Mentor should just advise the protégé but should not dictate. Entrepreneur E1P3 reported about that the mentor helped in avoiding the mistakes, because he had already gone through those phases that the entrepreneur was going through. Entrepreneur also reported that mentoring also provided motivational support and made the entrepreneur think he could achieve everything what his mentor had achieved. Entrepreneur E2P3 reported about the past mentoring experience where mentor helped in sales and assembly of computer hardware. It was the mentor who initiated the relationship and offered mentoring support, because mentor believed in the potential of the entrepreneur. Entrepreneur E2P3 also reported about another mentor who was very close to the entrepreneur, but later on relationship turned sour, because the mentor started to give instructions and did not let the entrepreneur to take decision on own. Entrepreneur believes that a mentor should give advice, but shouldn't dictate or impose the decision upon the entrepreneur.

*"I have a mentor I call him guru ji, I don't want to name him. He also has a mentor and he is pretty successful. I think everyone has it. Directly or indirectly they do have someone ideal or someone they are following and when they follow they try to improve things. My mentor did this why don't I improve things more. Or maybe they avoid those mistakes which their mentor has done"* E1P3.

*"When I did my bachelor degree I didn't have any mentor I did it alone except one of the computer shop guy. At that time I used to sale computer and get 250 margin. So the computer guy told me why you just sale computer and get 250. Come to us we will train you and you can*

*not only sale computer but also assemble and repair. You can earn more than and you can manage your own team. We will give you supply. He also became a very good mentor for me. I would like a person who notices me. Yeah this guy is ok why not I become his mentor. So that times Mr Yupp and his younger brother. They saw me and they actually offered me coaching and mentored me in sales and assembly. In return he became my supplier. So I'm very lucky that time"*

*"You know I mentioned you my mentor who is 18 year older than me. Now our relationship is kind of sour. What happened because he is very close to me he always advises so it's like he owns me? It is like he is giving instructions. It's like I am not having my own design. Because a good mentor will never dictate. Your decision is product of your own conclusion. You listen to everyone in the end you have to make decision" E2P3.*

Literature analysis suggests that past experience with mentoring plays an important role in shaping the intentions to be mentored again. There is little or no study which examines the influence of past experience with entrepreneurial mentoring to have another mentor in the future. In a study of mentoring intentions in large organizations by Leck, Orser, and Riding (2009), it was found that one of the main reason that employees didn't want to be mentored was due to their past experience. They saw little or no value in mentoring. Another reason for not being inclined towards mentoring was that employees felt that it was time for them to mentor other protégés and not being mentored by others. Carr and Sequeira (2007) identified that past experience with family businesses positively influencing the intentions towards entrepreneurship.

**6.7 Key Findings**

Key findings of the research can be summarized as follows. These findings can become the basis of future research on entrepreneurial mentoring. Table 3 presents the key findings and reference to the section in the empirical findings chapter for the interview excerpt.

**Table 3:** Key Findings

| Key Findings | Reference |
|---|---|
| It is not just entrepreneurial support that influences the entrepreneurs' intentions towards mentoring, but apart from the support, the type of the mentor is also important. | Entrepreneur E3P1, E4P1, E7P1, E5P2, E8P2 |
| Information supports (Marketing, Business, Sales) provided in higher education institutes is not applicable in the real world of entrepreneurship, there is a need of entrepreneurial academia which can provide hands on experience to the students. | Entrepreneur E14P1, E10P1, E11P1. |
| Factors related to surface level characteristics of the mentor have more influence on entrepreneurs' | Entrepreneur E8P1, E10P1, E12P1, E14P1, E4P2, E8P2, E3P3, E4P3 |

| intentions towards mentorship during the conception phase than start-up and growth phase. | |
|---|---|
| The religion of the mentor is an important factor that influences the nascent entrepreneurs' intentions towards mentorship. | Entrepreneur E8P1, E9P1, E13P1. |
| Entrepreneurs use online media (Social media, Internet) to get the support they need to become successful. | Entrepreneur E6P1, E3P2, E4P3 Section 4.3.3.1, 4.4.3.1, 4.5.3.1 |
| Personality and behaviour of the mentor towards the protégé can make or break the mentoring relationship. | Entrepreneur E10P1, E16P1, E15P1, E4P2, E1P3, E2P3. |
| Entrepreneurs at start-up and growth phase would like to have a mentor who can also work as a business partner. | Entrepreneur E4P2, E3P3, E5P3 |
| Entrepreneurs want a mentor to help the protégés in decision making but, they do not want mentors to impose the decision upon the entrepreneurs. | E2P2, E4P2, E5P2, E1P3, E2P3 |
| Growth phase entrepreneurs want more networking support from the mentor. They want to use business contacts of the mentor to get the projects. | Entrepreneur E3P3, E4P3. |
| Mentoring support provided within incubators is not effective, because the support provided by mentors is information support while entrepreneurs want knowledge and problem solving support. | Entrepreneur E1P2, E5P2 |
| Mentoring support provided by senior incubatees (start-up and growth phase) to junior incubatees (conception phase) can be more effective. | Entrepreneur E1P3, E3P3 |

## 7. Conclusion

Entrepreneur mentoring is being seen as one of the pillars in creating an entrepreneurial eco-system and increasing the survival rate of the new start-ups. Entrepreneurs need support (Entrepreneurial, Psychological and Role model) in order to take the right decision at the right time. A mentor can provide the necessary support to entrepreneur because a mentor has already been through all the stages, an entrepreneur going through.

Compatibility between the mentor and protégé is very important. It was explored that even if the mentor is a successful entrepreneur who can provide all the support that the entrepreneur needs, but if mentor's surface and deep level characteristics does not match with what the entrepreneur wants, then it is highly likely that entrepreneur will not go for that mentor or the relationship will not work in case the entrepreneur does.

Therefore, for a successful mentor and protégé relationship, it is important that the mentor is not only competent enough to provide all the support, but is also compatible with the choice of mentor that entrepreneur wants. Surface level characteristics of the mentor may not matter later in the relationship, but deep level characteristics, especially personality, experience and accessibility of the mentor does matter throughout the relationship.